# Fluctuating stripes at the onset of the pseudogap in the high-$T_c$ superconductor $Bi_2Sr_2CaCu_2O_{8+x}$

Colin V. Parker[1], Pegor Aynajian[1], Eduardo H. da Silva Neto[1], Aakash Pushp[1], Shimpei Ono[2], Jinsheng Wen[3], Zhijun Xu[3], Genda Gu[3], and Ali Yazdani[1]

[1]*Joseph Henry Laboratories & Department of Physics, Princeton University, Princeton, NJ 08544*

[2]*Central Research Institute of Electric Power Industry, Komae, Tokyo 201-8511, Japan*

[3]*Condensed Matter Physics and Materials Science, Brookhaven National Laboratory, Upton, NY 11973 USA*

**Doped Mott insulators have been shown to have a strong propensity to form patterns of holes and spins often referred to as stripes[1-5]. In copper-oxides, doping also gives rise to the pseudogap state[6], which transforms into a high temperature superconductor with sufficient doping or by reducing the temperature. A long standing question has been the interplay between pseudogap, which is generic to all hole-doped cuprates, and stripes, whose static form occurs in only one family of cuprates over a narrow range of the phase diagram[2,7]. Here we examine the spatial reorganization of electronic states with the onset of the pseudogap state at T* in the high-temperature superconductor $Bi_2Sr_2CaCu_2O_{8+x}$ using spectroscopic mapping with the scanning tunneling microscope (STM). We find that the onset of the pseudogap phase coincides with the appearance of electronic patterns that have the predicted characteristics of fluctuating stripes[8]. As expected, the stripe patterns are strongest when the hole concentration in the $CuO_2$ planes is close to 1/8 (per Cu)[2-5,8]. While demonstrating**

**that the fluctuating stripes emerge with the onset of the pseudogap state and occur over a large part of the cuprate phase diagram, our experiments indicate that they are a consequence of pseudogap behavior rather than its cause.**

Spectroscopic mapping with STM probes the spatial structure of electronic states, which reflects band structure effects[9] as well as the propensity of the system toward an ordering such as stripes or other charge ordered states[8,10-15]. While static order manifests itself as bias-independent modulations in the STM conductance maps[16],detection of fluctuating order is more difficult as it requires disentangling the effects of band structure from those associated with proximity to an ordered phase. Previous STM studies of $Bi_2Sr_2CaCu_2O_{8+x}$ have shown two types of features: those associated with the interference of quasi-particles from the superconducting state's band structure caused by scattering from impurities[9,15,17,18], and electronic modulations that could be due to charge organization[10,11,13,14,19]. However, there is a lack of evidence for any static order in the same compound from other experimental probes[20,21], but some experiments and theory suggest that fluctuating stripes may be responsible for the formation of the pseudogap[19,22-24]. Moreover, the relation of the STM features with the pseudogap, whether due to pairing, its fluctuations, or charge organization has remained unclear because experiments have been limited to temperatures below or just above the superconducting transition temperature, $T_c$[10-14,25]. Extending these experiments above the region of fluctuating superconductivity[26,27], $T_o$, up to the onset of the pseudogap[6] at T* has been technically challenging.

Figure 1 shows examples of real space low-energy conductance mapping (G(r,V) = dI/dV(r,V), see supplementary section A) with the STM on an underdoped $Bi_2Sr_2(Ca,Dy)Cu_2O_{8+x}$ sample (UD35) carried out at three different temperatures corresponding to the superconducting state at $T < T_c$, the fluctuating paired state with weak diamagnetic response[26,27] $T_c < T < T_o$, and the pseudogap state at $T_o < T < T^*$. Discrete Fourier transforms (DFTs) of the conductance maps display strong peaks at wavevectors marked Q* (Fig. 1d-f) along the Cu-O bond direction in all three temperature ranges. Similar strong features appear at a larger wavevector, marked Q**, also along the Cu-O bond direction. At temperatures below $T_c$, the DFTs of the conductance maps show additional peaks that have been previously associated with the interference of Bogoliubov-de Gennes quasi-particles (BdG-QPI) in the superconducting state[9,15,17,25]. The contribution to the spatial variation of the density of states with superconducting origin can be further enhanced by the use of a ratio of conductance maps[17] $Z(r,V) = G(r,+V)/G(r,-V)$ (due to particle-hole symmetry), measured at the same energy, as demonstrated in Figures 1g-i (peaks marked q2-q7).

All the modulations in the density of states in our measurements, including those corresponding to Q* and Q**, have energy-dependent wavelengths near the Fermi energy (±50 mV) throughout the entire temperature range and hence are not due to the formation of static charge ordering (see supplementary section B). However, a distinction between Q* and Q** and the other wavevectors that occur due to impurity-induced interference becomes evident when analyzing their properties across a range of energies. It has been proposed that signatures of incipient order appear in STM conductance maps with similar phases across a range of energies, whereas modulations of the density of states due to impurity-induced quasi-particle interference are phase incoherent[8,12]. To distinguish which features are associated with incipient order, we compute the quantity $S(k_x, k_y, V_0) = \left| \int_{-V_0}^{V_0} G(k_x, k_y, V) dV \right| \Big/ \int_{-V_0}^{V_0} \left| G(k_x, k_y, V) \right| dV$, where $G(k_x, k_y, V)$ are the DFTs of conductance maps measured at different energies $V$, and $V_0$ defines the energy range over which we examine the phase of the modulations. This quantity measures the phase matching of modulations between different energies, and is independent of whether or not the features are dispersive (see supplementary section C).

Figure 2 shows that integration over a range of energies near the Fermi level results in the suppression of the wavevectors labeled q2-q7, indicating that they are due to impurity-induced interference phenomena. In contrast, the Q* and Q** features are not suppressed by this integration, even above $T_c$ and $T_o$ (Figure 2c). In figure 2d, we also show the detailed behavior of all the wavevectors in S with increasing integration window $V_0$. This analysis establishes the phase coherent properties of the Q* and Q** modulations and shows that their behavior is consistent with that expected from an incipient charge organization phenomenon. While there is the possibility that the Fermi surface of $Bi_2Sr_2CaCu_2O_{8+x}$ gives rise to impurity-induced bond-oriented interference wavevectors near Q* and Q**[17,28], the analysis above establishes that the majority of the contribution to our signal near these wavevectors is associated with an incipient order. The intensity of Q* versus energy also shows a strong enhancement near a specific wavevector, which is another predicted feature of incipient order (see supplementary section B). Although there have been indications from previous studies[8,10,11,14] that Q* and Q** modulations are not simply due to quasi-particle interference phenomena, the measurements and analysis presented here firmly establish that these modulations are due not to static but to fluctuating order.

The Q* and Q** wavevectors behave similarly in the S maps as shown in Figure 2b-d; however, their energy dependence is different. Q* dominates at lower energies (below 50 mV) while Q** is strongest at higher energies (see supplementary section D). Since only an incipient order that influences electrons near the Fermi level is relevant to the low energy properties of our system, such as its transport properties, we focus our attention on the modulations at Q*.

We find further distinction between the fluctuating order-induced Q* and the impurity-induced wavevectors, which allows us to establish separate connections to different regions of the cuprate phase diagram when we study their temperature dependence. As evidenced in Figure 1, the BdG-QPI wavevectors (labeled $q_2$-$q_7$) are strongly suppressed above $T_c$, while Q* remains robust at all temperatures below T*. Examining conductance maps over the entire phase diagram, we have measured the power spectral density (PSD) of Q* at 10 meV (see supplementary section D) over a wide range of temperature and doping. Figure 3a shows that the onset of Q* coincides with the onset of pseudogap behavior at T*, the value of which has been determined for our samples by temperature dependent spectroscopic measurements of pseudogap disappearance[29]. In strong contrast, the intensity (PSD) of superconducting BdG-QPI modulations across the phase diagram, such as $q_7$ shown in Figure 3b, is strongly suppressed when the superconducting phase coherence in the sample is lost above $T_c$. This behavior is consistent with recent theoretical modeling of BdG-QPI as a function of temperature[30].

In contrast to the pseudogap, we find that the intensity of Q* has a non-monotonic doping dependence (Figure 3a). Plotting the intensity of Q* in DFTs at an intermediate temperature as a function of doping (Figure 3c), we find a remarkable peak in the intensity of Q* when the nominal hole concentration of the sample is close to 1/8. In contrast, the intensity of $q_7$ appears to peak near optimal doping. The 1/8 hole concentration corresponds to the ideal doping for formation of half-filled stripes, as has been well established by neutron scattering experiments on La-based compounds[2] and in model calculations of cuprate properties[1,4,5,8]. The unique doping dependence shown in Figure 3c suggests that Fermi surface nesting near the anti-node cannot be responsible for the robust Q* but rather identifies fluctuating half-filled stripes as the origin of these modulations. Moreover, since the $Bi_2Sr_2CaCu_2O_{8+x}$ system does not undergo a structural distortion near 1/8 doping, we conclude that local Mott and anti-ferromagnetic correlations, as opposed to structural distortion as in the La-based systems, are stabilizing fluctuating stripe patterns near this hole concentration. Nevertheless, distinguishing whether the observed patterns are strictly one dimensional (fluctuating stripes) or two dimensional (checkerboards) is complicated by the presence of doping inhomogeneities which locally pin these modulations [31].

The connection between the appearance of the incipient fluctuating stripe order and T* is shown in figure 4. We contrast real space STM conductance maps together with spatial maps of the pseudogap energy measured at the same atomic location at a temperature above $T_c$ near optimal doping. At this temperature and above, consistent with our previous studies, we find regions of the sample in which the pseudogap has collapsed[29]. Such spatially inhomogeneous suppression of the pseudogap is strongly correlated with the disappearance of Q*. This behavior is evident when comparing the conductance map in Figure 4a with the pseudogap map in Figure 4b. Clearly, the regions that show modulations also exhibit the strongest pseudogap. This can also be seen in Figure 4c, which plots a local measure of the modulation strength related to smoothed and normalized products of the data and sinusoidal functions, a real space intensity map of Q* (see supplementary section E). It is clear that regions with enhanced pseudogap nucleate Q* modulations at higher temperatures.

Examining the strength of incipient fluctuating stripe order and pseudogap phenomena as a function of doping points to disparate behavior between these two important phenomena and provides important clues on their connection. As shown in Figure 3c, below 1/8 doping, the strength of fluctuating stripes is diminished. Furthermore, examining the local correlation between the Q* modulations and the pseudogap, we find that the cross-correlation between these two phenomena switches sign from positive to negative near the nominal doping of 1/8 (Fig. 4d). Approaching the Mott insulating state by lowering the doping results in a stronger pseudogap, while propensity to stripe formation is suppressed. Considering these facts together, we conclude that the pseudogap is not caused by stripe correlations, but rather that the pseudogap is required to nucleate fluctuating stripes. Given that stripe formation is believed to occur through phase separation of mobile holes and anti-ferromagnetically correlated regions, it seems logical that the pseudogap is associated with local spin correlations that make the formation of such spatial patterns possible.

We gratefully acknowledge discussions with A. N. Pasupathy, P. W. Anderson, D. Huse, S. Kivelson, E. Fradkin, and, N. P. Ong. This work was primarily supported by grant from DOE-BES. The instrumentation and infrastructure at the Princeton Nanoscale Microscopy Laboratory are also supported by grants from NSF-DMR, NSF-MRSEC programme through Princeton Centre for Complex Materials, and the W.M. Keck foundation.

C. P., E. H. S. N., P. A., and A. P. performed the STM measurements; C. P. and P. A. analysed the STM data; S. O., J. W., Z. X., and G. G. prepared the crystals; A. Y. supervised; and A. Y., C. P., E. H. S. N., and P. A. wrote the manuscript.

The authors declare no competing financial interests.

Correspondence and requests for materials should be addressed to A.Y. (yazdani@princeton.edu).

**Figure captions:**

**Fig.1: Electronic modulations in $Bi_2Sr_2(Ca,Dy)Cu_2O_{8+x}$.** Real space low-energy conductance maps G(r,V) at V=10 mV on UD35 ($T_c$=35 K) carried out at 8 K (a), 41 K (b), and 122 K (c). At all temperatures, strong electronic bond-oriented modulations are observed. The white scale bars in (a,b,c) correspond to 5 nm. The corresponding DFTs of G(r,V=10 mV) and of the Z(r,V=10 mV) are respectively shown in (d,e,f) and (g,h,i), which display strong peaks at several wavevectors as marked in (g). Only the peaks marked as Q* and Q**, corresponding to the real space modulations in (a,b,c), survive to temperatures well above $T_c$. The DFTs in (d-i) are normalized to their corresponding standard deviations.

**Fig.2: Impurity induced interference versus incipient order**. (a) DFT of the Z-map at 10 mV on UD35 carried out at 8 K (normalized by its standard deviation as in figure 1). The circles show all the wavevectors (Q*, Q**, and q2-q7) that show pronounced peaks. (b) S-map of the same dataset as in (a) integrated within $\pm V_0$=25 mV (see main text). Only the modulations at Q* and Q** survive this integration procedure. (c) Same dataset as in (b) carried out at 70 K displaying similar behavior. (d) Relative S-map intensity of the different wavevectors as marked in (b) as a function of the integration window $V_0$. The position of the different wavevectors is determined by the peak locations in the Z map . The dashed grey line represents the decay profile of the background

**Fig.3: Phase diagrams.** (a) Intensity (PSD) of the Q* modulation over the entire temperature-doping phase diagram for $Bi_2Sr_2CaCu_2O_{8+x}$. The intensity comes from the amplitude of Gaussian fits to the peak, and is plotted on a logarithmic

scale, and in many cases represents the average taken from multiple measurements on different samples from the same batch of crystals. The shaded grey area, the red solid line, and the dashed green line represent the pseudogap T* (determined from previously reported spectroscopic measurements of pseudogap disappearance on similar samples[29]), fluctuating superconductivity $T_o$, and the superconducting, $T_c$, phases respectively. (b) Intensity of the $q_7$ modulation over the same temperature-doping phase diagram. (c) Intensity of both modulations (Q* and $q_7$) near 35 K showing that the intensity of Q* is maximal near 1/8 hole doping whereas that of $q_7$ peaks near optimal doping. The systematic error in Q* intensity is 2 dB.

**Fig.4: Spatial correlation of the Q* modulations with the pseudogap.** (a) Real space STM conductance map G(r,V=10 mV) of an OP91 ($T_c$=91 K) sample at T=102 K. The map shows the spatially inhomogeneous disappearance of the Q* modulations which mirror the inhomogeneous suppression of the pseudogap shown in (b). (c) Local measure of the modulation strength (see supplementary section E), which shows that regions with strong modulations also exhibit large pseudogaps. (d) Measure of the local correlation of the Q* modulations with the pseudogap for different dopings showing that the correlation switches sign near 1/8 doping. The error bars indicate the amount of anomalous correlation that would be expected in datasets of equivalent size.

# Figure 1

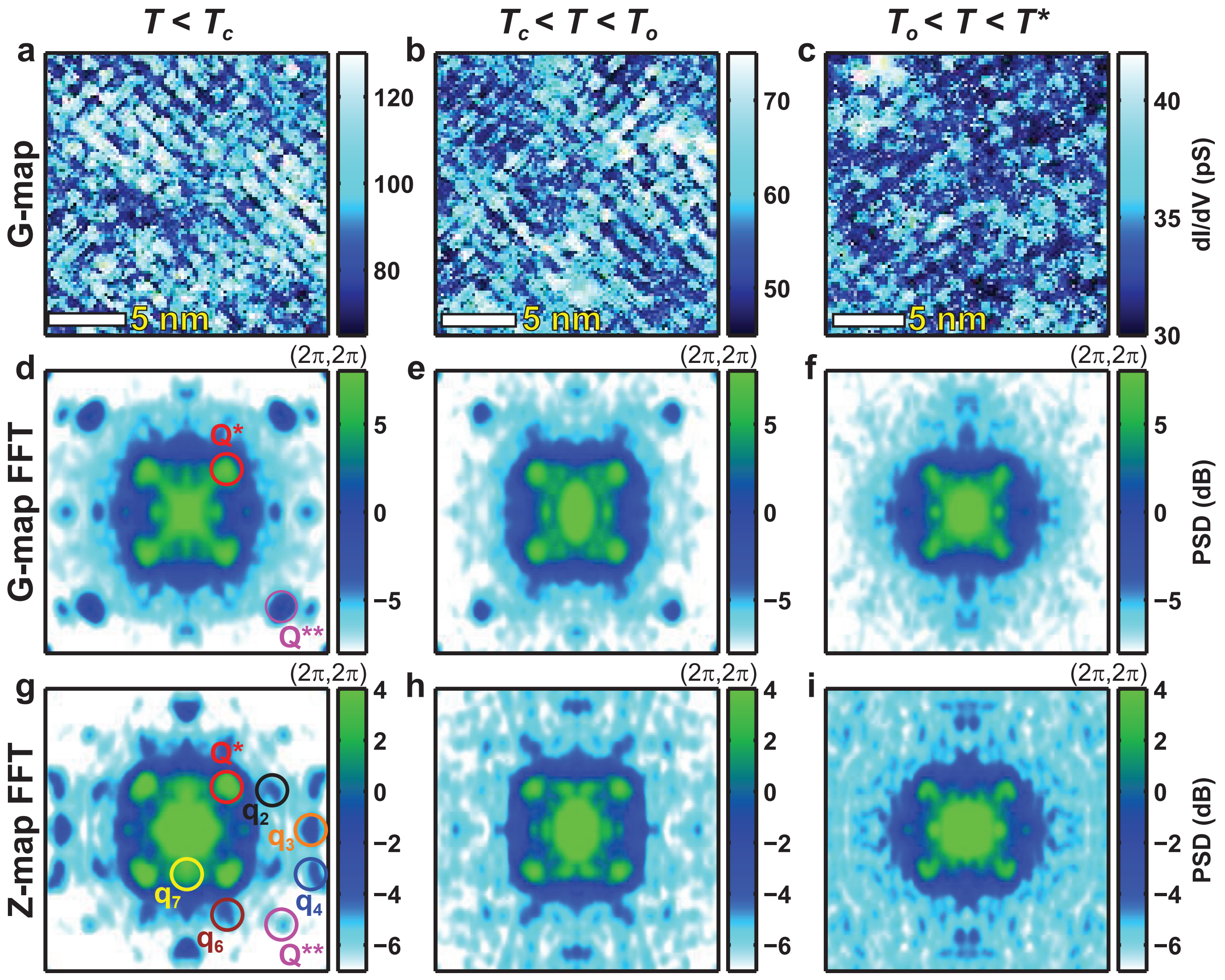

# Figure 2

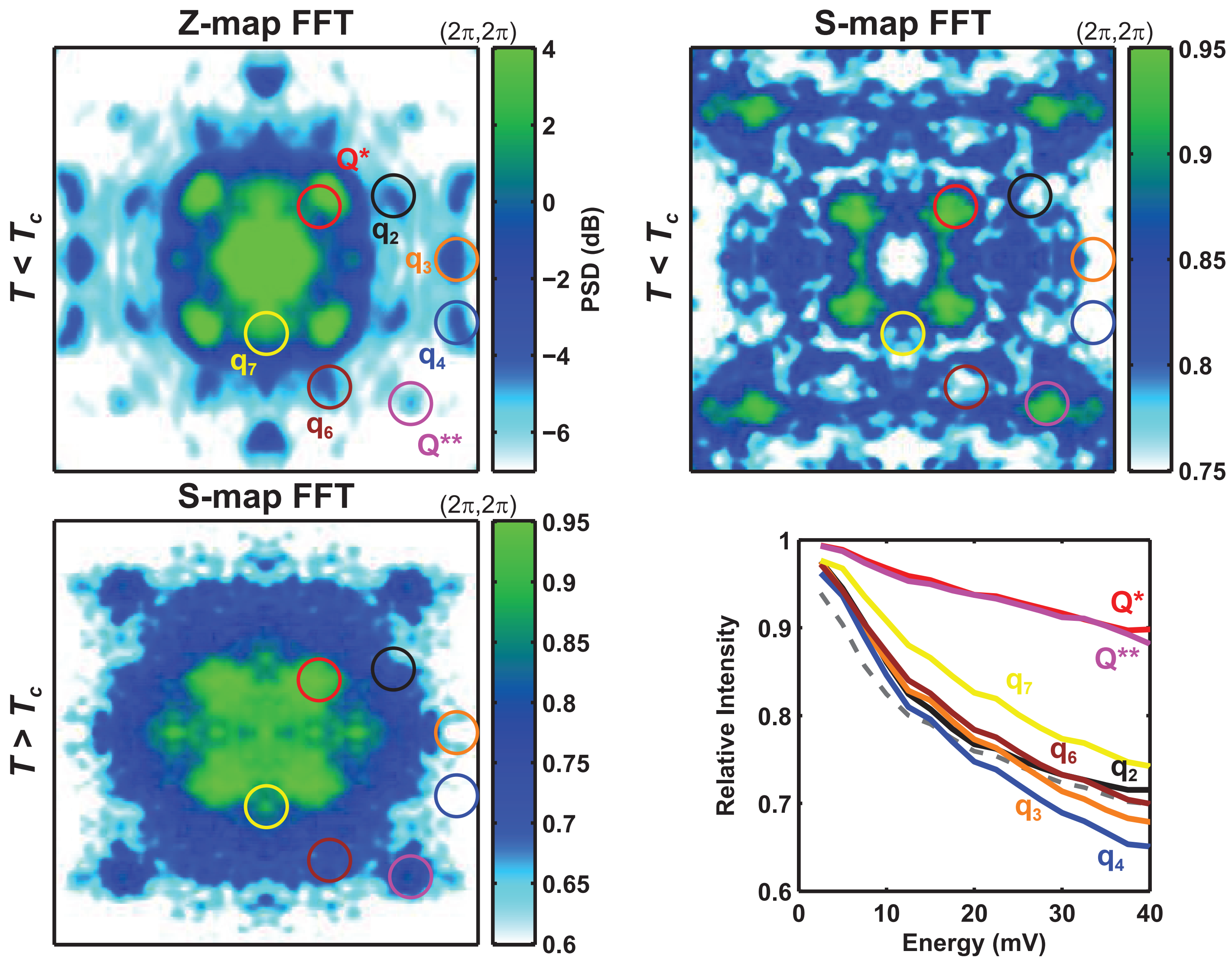

# Figure 3

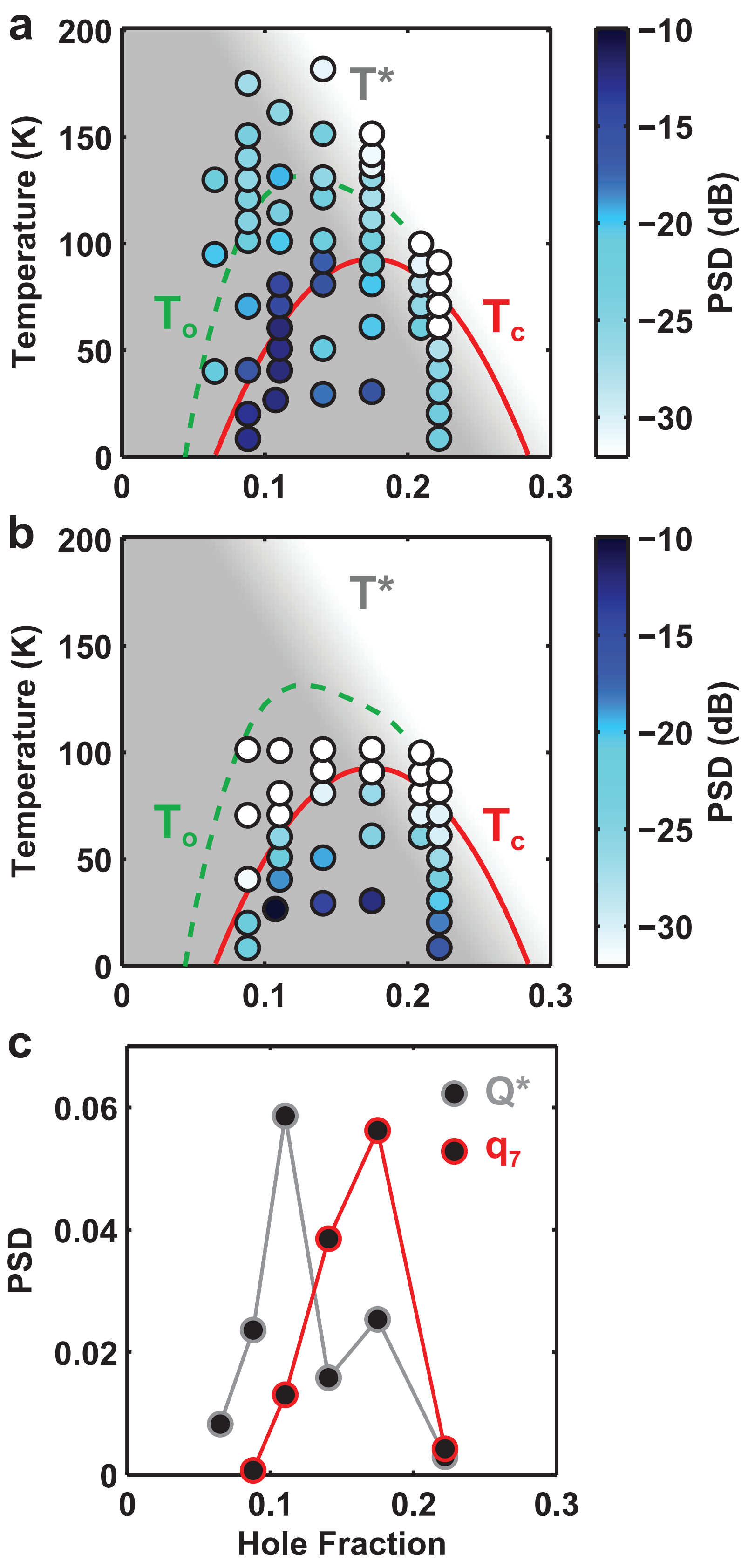

# Figure 4

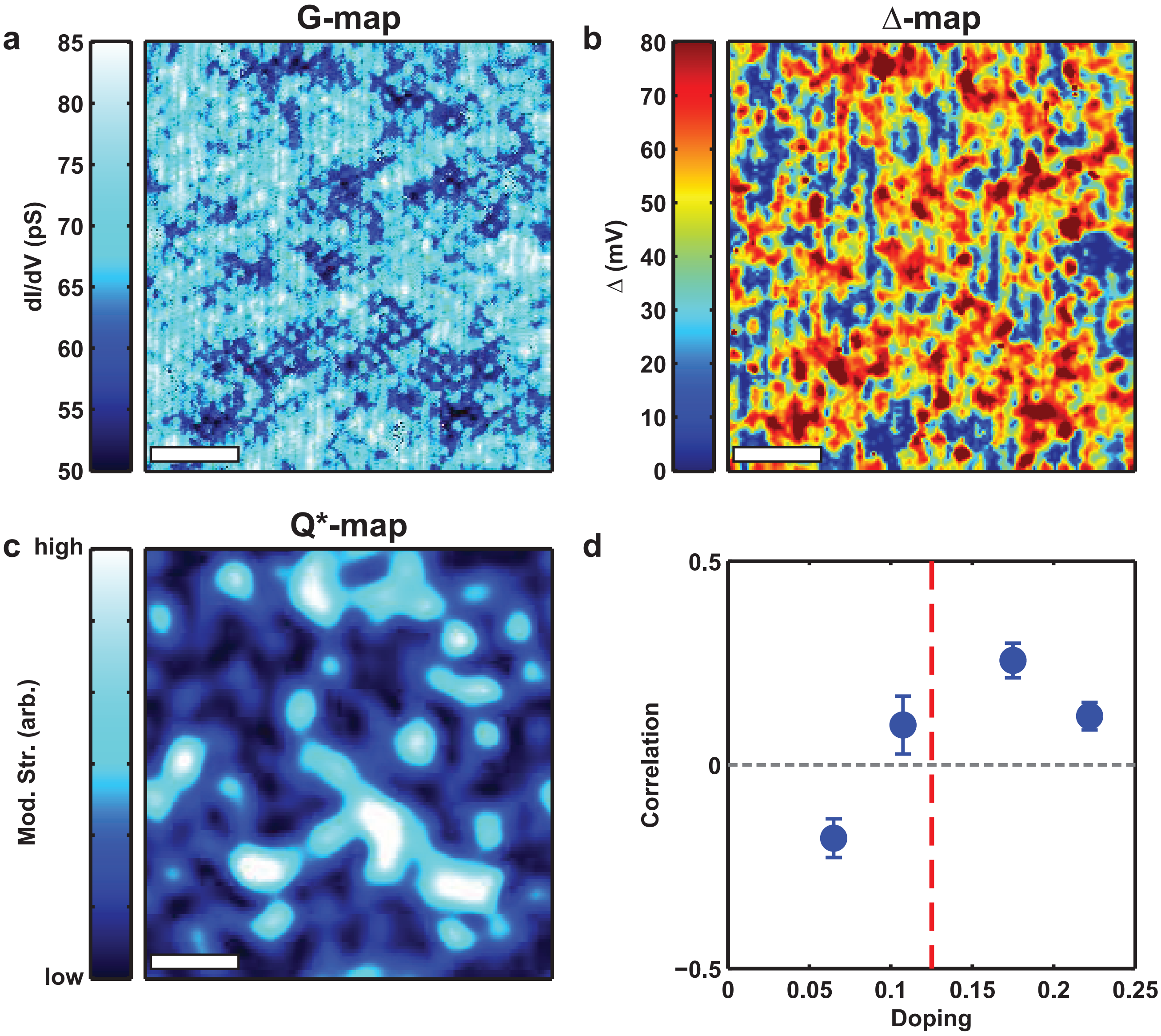

## Supplementary Information

Colin V. Parker, Pegor Aynajian, Eduardo H. da Silva Neto, Aakash Pushp, Shimpei Ono, Jinsheng Wen, Zhijun Xu, Genda Gu, and Ali Yazdani

### A - Data Acquisition & Analysis

The data are taken on a homebuilt variable temperature ultra-high vacuum scanning tunneling microscope. The conductance maps are measured using standard lock-in techniques, and always using the same setpoint bias of -150 mV. We typically choose a modulation level near $k_B T$ (8.6 mV for 100 K) and a setpoint current between 20 and 80 pA. To reduce the impact of noise due to motion of the atoms at the end of the tip at high temperature on our data, we utilized a procedure to reject noise due to these instabilities. We reject signals in the conductance maps that are two standard deviations away from the average of the map. At these pixels, we use an average value of the nearby pixels through a simple smoothing procedure before we carry out further analysis.  After carrying out the discrete Fourier transforms (DFT), we use the mirror symmetry (relative to the a-axis of the crystal) to further suppress influence of random tip noise on our measurements.

For the purpose of comparing the intensity of features between different measurements, we follow a simple procedure that takes into account whether the maps were obtained at different junction impedances or over differently sized regions of the samples. We first normalize the maps by their mean prior to performing the DFT, to eliminate any dependence on junction impedance. We account for different area sizes by scaling the results of the DFT by the appropriate factor, plotting the results in units of power spectral density times Cu-O plaquette area.

To obtain peak location and intensity of feature in DFT maps, we must accurately remove the impact of the central peak (due to sample inhomogeneity). An example of this is shown in Figure S1. By taking multiple cuts through the peak of interest in the DFT (such as Q* in Figure S1) and the regions that do not included this peak, we isolate the signal from the background. The resulting signal is fitted with a Gaussian. The intensity, or power spectral density, is the amplitude of the fitted Gaussian.

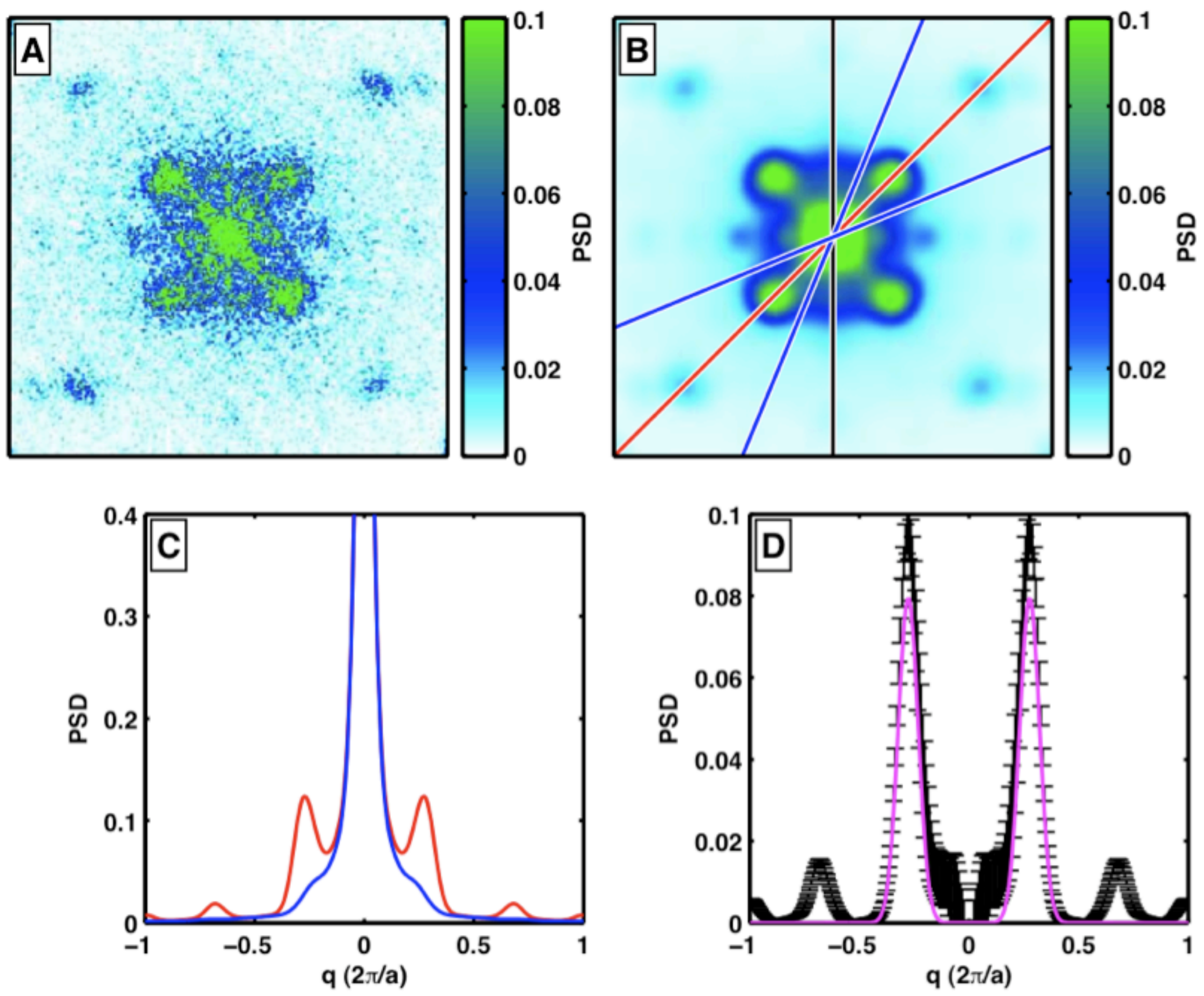

**Figure S1** – Peak fitting procedure. **(A)** Raw DFT dI/dV measurement from an underdoped sample with $T_c$ = 35 K taken at 8 K. **(B)** The same measurement after being smoothed **(C)** Line cut profiles of the data in **B**, along contours through the peak (red) and beside the peak (blue). **(D)** Difference between the peak line cut and the background in **C**, (black point with error bars). The magenta line shows a Gaussian fit used to extract the position and intensity of the peaks.

## B - Dispersion of the Modulations

The Q* modulations have a bias dependence both above and below $T_c$, and hence cannot be coming from a static order. Figure S2 shows the fitted peak position for Q* as a function of bias for three different samples taken at several temperatures both above and below $T_c$. For the optimally doped sample, below $T_c$, there is a clear symmetric dispersion, which is expected for a particle-hole symmetric superconducting state. However, as indicated by the black line in Figure S2C, the detailed predictions of the "octet model"[31] based on the band structure calculation by Norman[32] do not agree well with the dispersion. Above $T_c$ for optimal doping, the particle-hole symmetric dispersion is lost, but the wavevector is changing with bias nonetheless. However, while the symmetric dispersion can also be seen in the dispersion of the Z-ratio, the asymmetric dispersion is not apparent in the Z-ratio dispersion, since Z-ratio implicitly suppresses an anti-symmetric component to the dispersion. For the lowest doping shown (UD35), even the dispersion below $T_c$ is not symmetric, though there is dispersion at all temperatures. The absence of particle-hole symmetry is also an indication that the Q* modulations are not the result of scattering in a phase-incoherent superconductor.

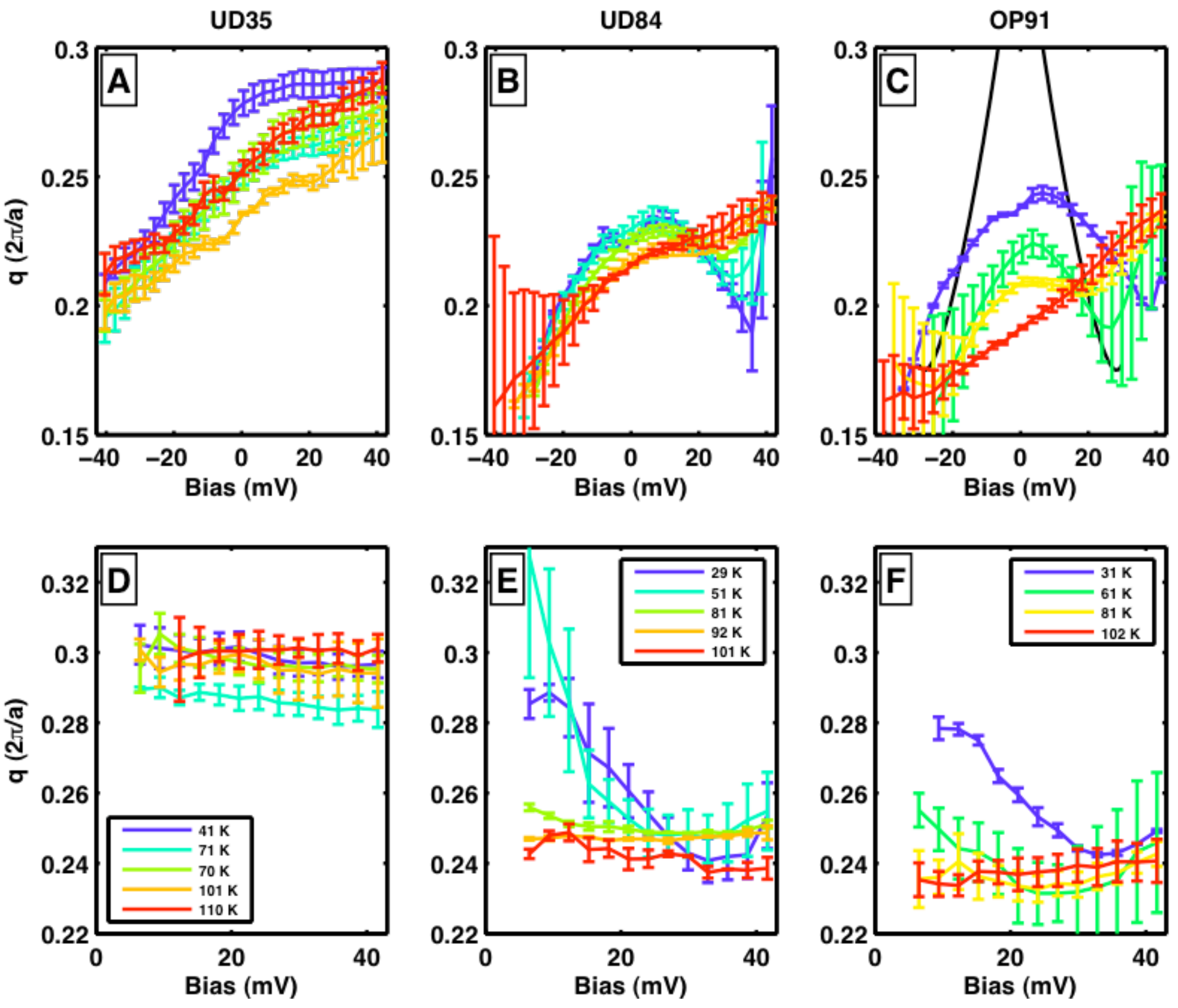

**Figure S2** – Dispersion of Q* modulations .**(A-C)** Bias dependence of the Q* modulation wavevector for different samples. At low temperatures, a symmetric component is visble, which goes away near T*, leaving an asymmetrically dispersing background. The black line in panel C illustrates the behavior expected from impurity scattering of the superconducting state[1,2]. **(D-F)** The same dispersion taken from Z-ratio maps. In this case, the symmetric portion of the dispersion can be seen, but the asymmetric portion does not show up.

Another indication that Q* is not due to static stripes or BdG-QPI can be found by plotting the Q*'s modulation strength (PSD) against the observed wavevector, as shown in figure S3. From this plot it is apparent that the intensity peaks near a particular wavelength, not far from the commensurate 1/4 value expected by many stripe models. The dispersion and the intensity of Q* are affected by both the stripe wavevectors and the bandstructure and requires a model that includes both effects.

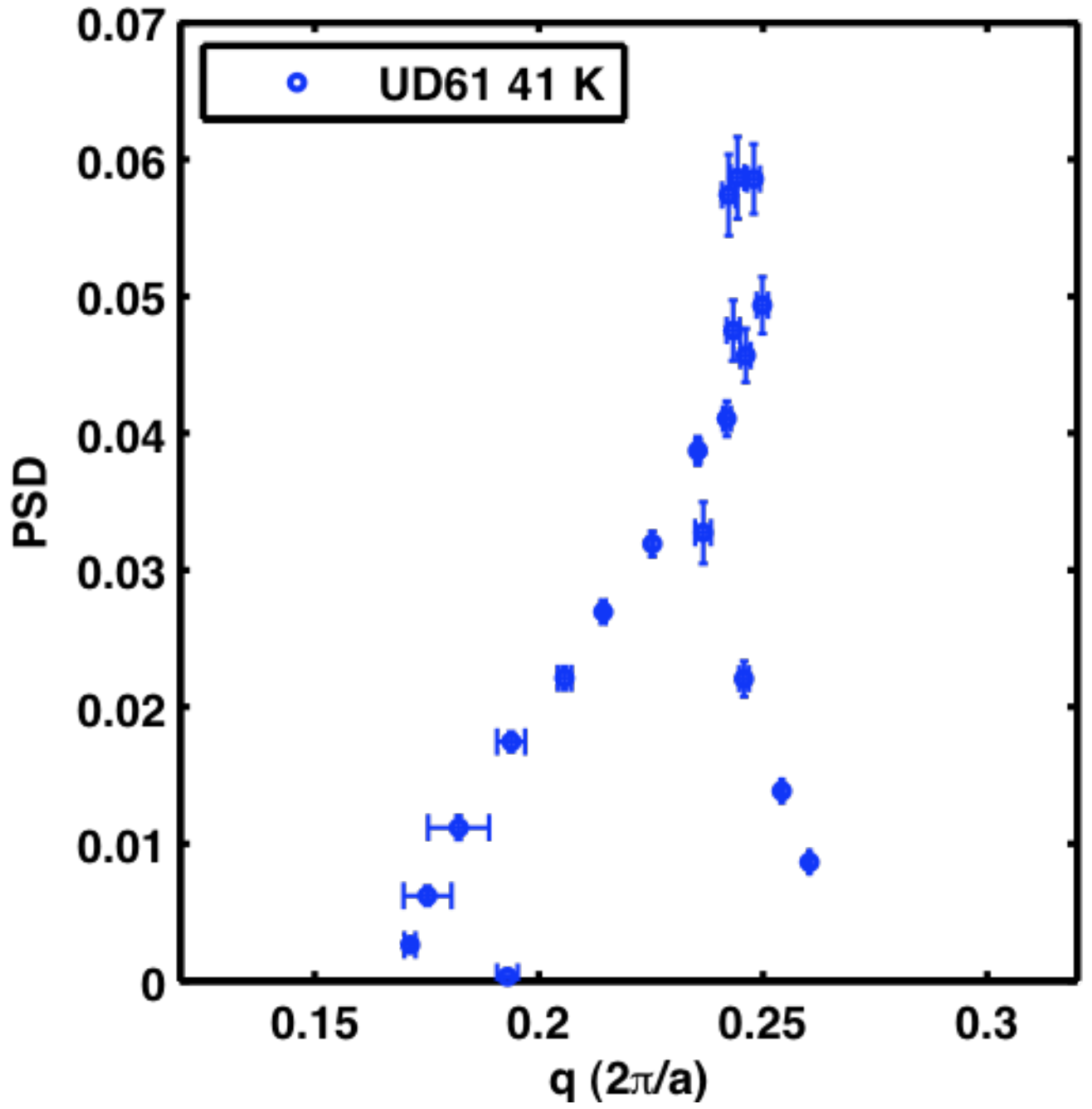

**Figure S3** – Wavevector versus intensity for Q* modulations, taken from sample UD61 at 41 K. Note the peak in intensity near q = 0.25.

**C – Phase matching and the S map**

As stated in the paper, whether features appear or not in the S map is independent of their dispersion. This is because the S map concerns the ratio of two quantities that differ only in their treatment of the *phase* of the modulations. It is possible in fact for even a non-dispersive feature to be suppressed in the S map, if the phase (but not the wavelength) changes with energy. The S map will tend to suppress dispersive features, unless as the feature disperses through a given point in k-space, the leading edge at one bias has the same phase as the trailing edge at a different bias. This is uniquely the case for Q* and Q** modulations and not for the other peaks.

To illustrate the importance of the phase in S and that summation of maps does not artificially suppress dispersing features regardless of the phase coherence, we show the two integrals whose ratio leads to the S map as separate quantities. Figure S4A shows a sum of conductance maps, where we take the absolute value before we sum them. This panel shows many features, including Q*, Q** and features associated with QPI, like q2, and q3 (compare to figure 2a). The presence of these features shows that even strongly dispersing peaks do not disappear into the background in S. In figure S4B, we repeat the procedure on the same data set, but we take the absolute value after summing over the same voltage range as in figure S4A (the S-map in the paper results from dividing figure S4B by figure S4A). In figure S4B, the features associated with QPI are suppressed because their phase is sufficiently different over the energy window of the analysis. The phase-coherence property of Q* and Q** in real space are what allows them to survive the summation in figure S4B.

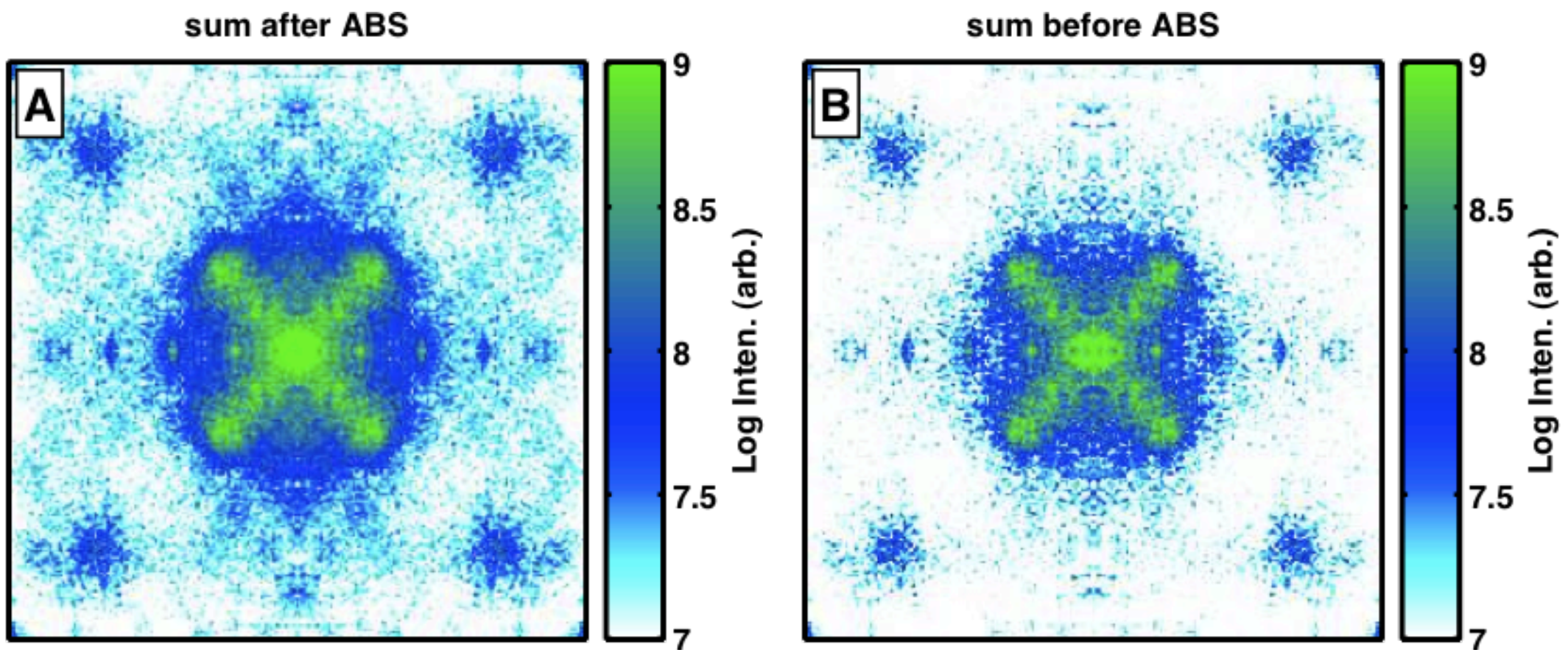

**Figure S4** – Individual components whose ratio leads to the S map. **(A)** The denominator of the S map, the sum of all DFTs with the absolute value taken before the sum. **(B)** The numerator of the S map, the sum of all DFTs with the absolute value taken after the sum.

### D - High Bias and Fine Structure

Though both Q* and Q** show an energy dependence that cannot be from impurity-induced scattering, they have very different behavior over large bias ranges. As shown in Figure S5, Q* modulations are most intense near the Fermi energy, and are almost gone by 50 mV. In contrast, Q** is most intense at higher energies near the pseudogap energy[33]. This difference makes measurements of Q** more challanging than Q*, as they require that the STM bias point to be taken into account. This can be accomplished by either measurements with very high bias or analyses of quantities that are independent of the set point[34].

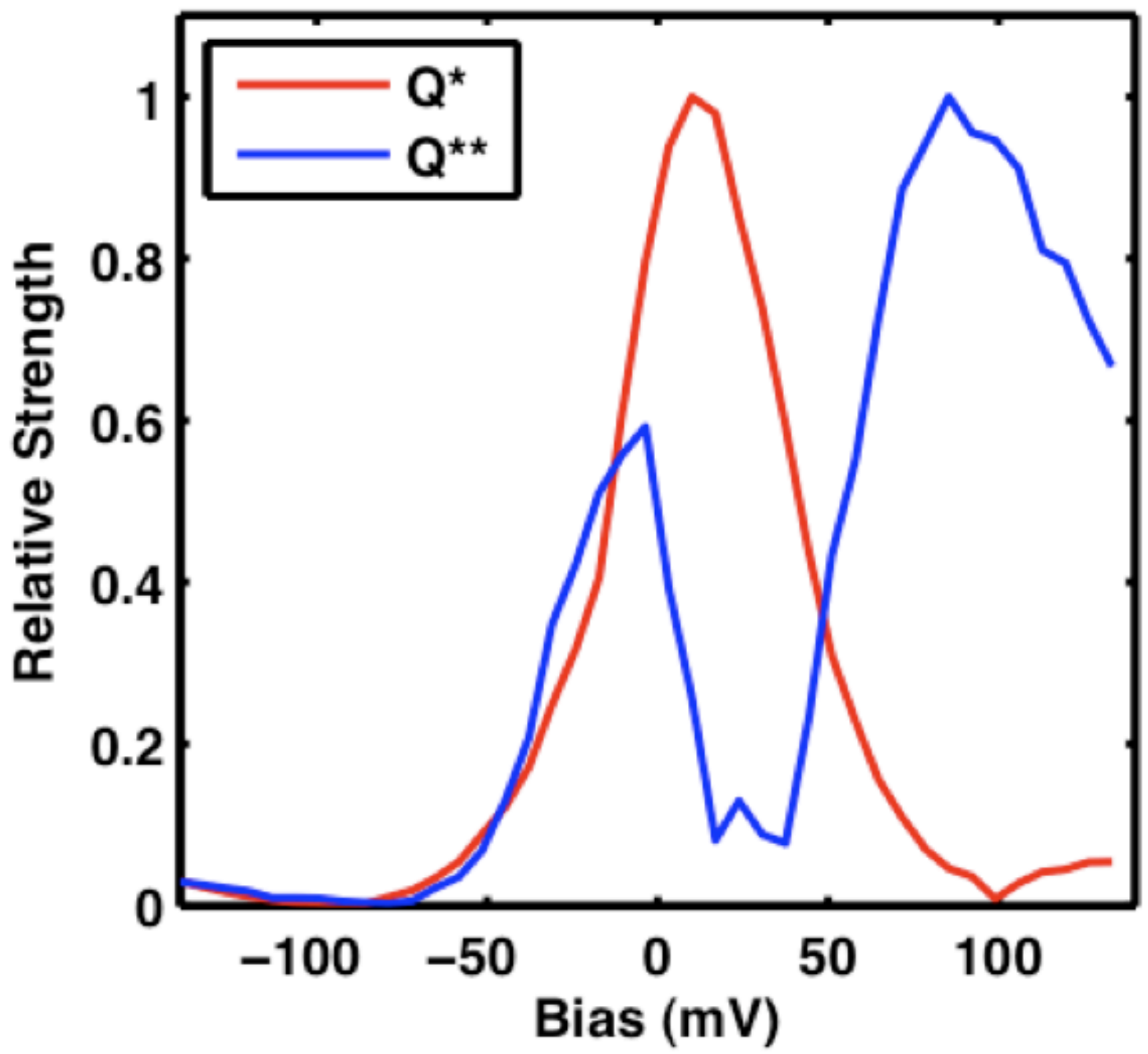

**Figure S5** – High bias and fine structure. The normalized intensity of the modulations at Q* and Q** taken from the underdoped sample with $T_c$ = 35 K at 20 K, with the STM setpoint at 150 mV. Q* has predominantly low energy contributions, and so is largely independent of setpoint above 100 mV.

### E - Measuring Q* locally

In order to determine the local strength of the Q* modulations, we first determine the location of the Q* modulation peak in the DFT. Next, we produce simulated modulations consisting of a pure sine and cosine at exactly the same wavevector as the Q* peak. We multiply this point by point with the data in real space. We then smooth in order to remove the 2Q* component. Finally, we divide by a smoothed version of the original data set. This last step ensures that we measure the local strength of the modulation relative to the background conductance. This process done separately for sine and cosine contributions, and for each of the Q* peaks (a + b direction, and a - b direction). The results are added in quadrature to give Figure 4c of the main paper.

### F – Error Quantification

For quantities derived from fits to the peaks, we can use the fitting algorithm to also determine error bars based on the noise level of the measurements. This is done in

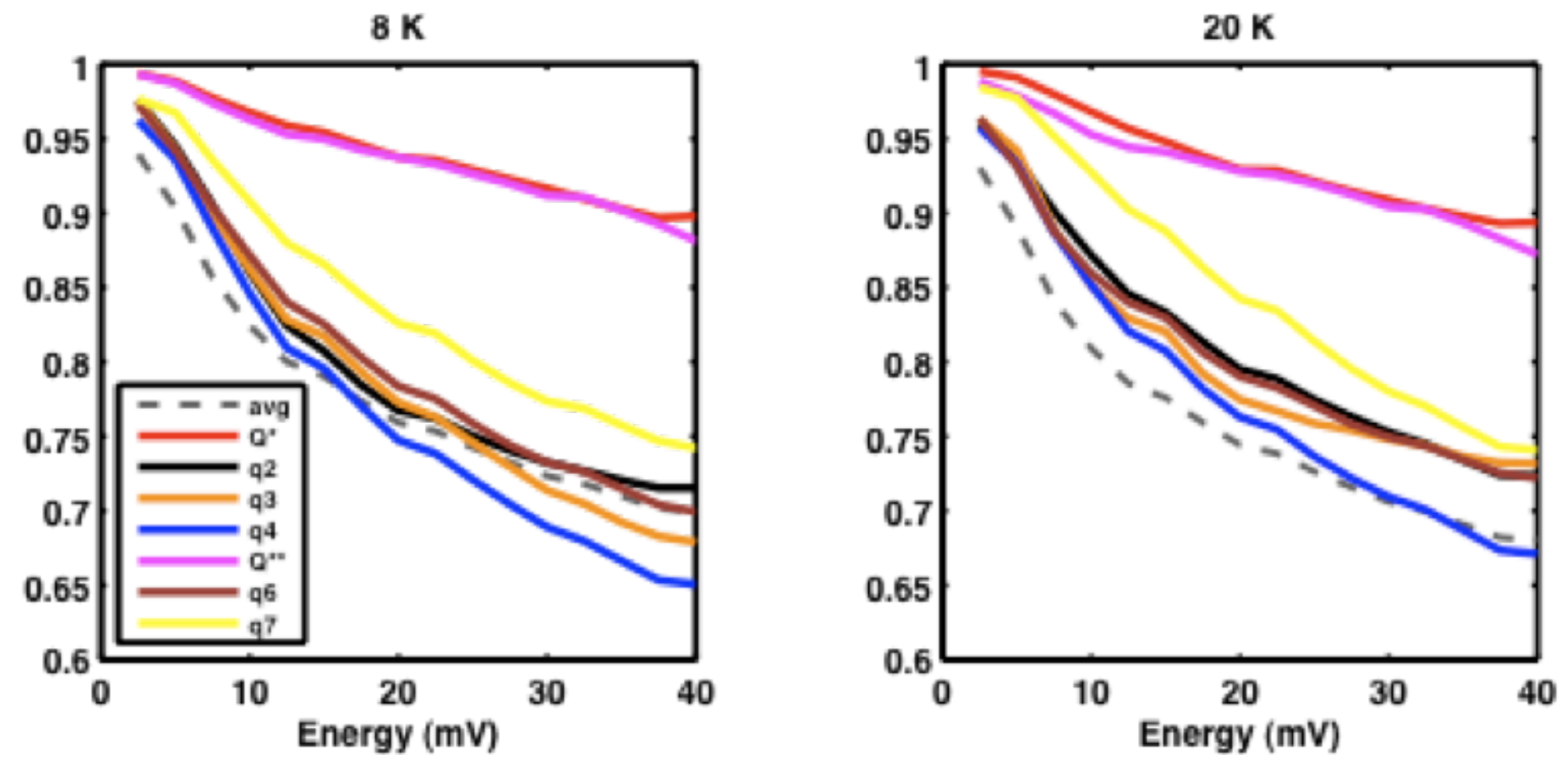

**Figure S6** – Comparison of figure 2d from the main text for two different temperatures on the same sample (UD35). Because of the strong similarity we conclude that the separation of Q* and Q** from the other peaks is well above any experimental error.

supplementary figure S2. However, a much larger source of error is the variation of the observed modulation strength between different STM junctions (caused by different tip configurations). To understand the impact of these variations on our results, we have performed four measurements at the same temperature (50 K) on the same sample (UD61) with different junctions. The standard deviation for the power spectral density was 2 dB, which is 10% of the full-scale variation in figure 3.

To estimate the error in figure 2d, we have repeated the analysis for a different junction on the same sample (UD35) at a different temperature (20 K vs 8 K) (see figure S6). The near identical behavior between the two measurements shows that the separation of Q* and Q** from the other peaks is much larger than experimental error.

The sample doping is determined from measurements of its $T_c$, and then connected to doping measurements[35]. Based on figure 6 from the above reference it would appear that doping error is in the range of 2% hole concentration.

## Supplementary References